\documentclass{article}
\usepackage{cite}
\usepackage{graphicx}
\usepackage{dcolumn}

\begin{document}

\date{}
\title{Comment on: ``On the influence of a Coulomb-like potential induced by the
Lorentz symmetry breaking effects on the harmonic oscillator''. Eur. Phys.
J. Plus (2012) \textbf{127}: 102}
\author{Francisco M. Fern\'{a}ndez\thanks{%
fernande@quimica.unlp.edu.ar} \\
INIFTA, DQT, Sucursal 4, C.C 16, \\
1900 La Plata, Argentina}
\maketitle

\begin{abstract}
We analyze the calculation of bound states for a nonrelativistic spin-half
neutral particle under the influence of a Coulomb-like potential induced by
Lorentz symmetry breaking effects. We show that the truncation condition
proposed by the authors only provides one energy eigenvalue for a particular
model potential and misses all the other bound-state energies. The
dependence of the cyclotron frequency on the quantum numbers is a mere
artifact of the truncation condition that is by no means necessary for the
existence of bound states.
\end{abstract}

Some time ago Bakke and Belich\cite{BB12} obtained the bound states for a
nonrelativistic spin-half neutral particle under the influence of a
Coulomb-like potential induced by the Lorentz symmetry breaking effects.
They claim to present a new possible scenario of studying the Lorentz
symmetry breaking effects on a nonrelativistic quantum system defined by a
fixed space-like vector field parallel to the radial direction interacting
with a uniform magnetic field along the $z$-axis. They also discuss the
influence of a Coulomb-like potential induced by Lorentz symmetry violation
effects on a two-dimensional harmonic oscillator. Starting from a
time-dependent Schr\"{o}dinger-Pauli equation, and by means of a series of
transformations the authors derive a three-term recurrence relation for the
coefficients of the power-series approximation to the radial function in
cylindrical coordinates. By suitable truncation of the recurrence relation
they obtain an analytical expression for the energies and a surprising
dependence of the oscillator angular frequency on the quantum numbers. They
interpret this result in the following way: ``we have shown that Lorentz
symmetry breaking effects do not break the degeneracy of the energy levels
of the harmonic oscillator, but impose a dependence of the cyclotron
frequency on the quantum numbers $n$, $l$ and $s$.''

The purpose of this Comment is the analysis of the approach applied by Bakke
and Belich\cite{BB12} to the quantum-mechanical equations derived from the
proposed models and in particular the truncation method used by them and its
effect on the conclusions drawn in their paper.

The starting point of present discussion is the eigenvalue equation for the
radial part of the Schr\"{o}dinger equation
\begin{eqnarray}
&&G_{s}^{\prime \prime }+\frac{1}{\xi }G_{s}^{\prime }-\frac{\nu _{s}^{2}}{%
\xi ^{2}}G_{s}-\frac{\beta }{\xi }G_{s}-\xi ^{2}G_{s}+WG_{s}=0,  \nonumber \\
&&\nu _{s}=l+\frac{1}{2}(1-s),\;\beta =\frac{\delta }{\sqrt{m\omega }},\;W=%
\frac{\zeta ^{2}}{m\omega },  \nonumber \\
&&\delta =2gbB_{0}\nu _{s}+sgbB_{0},\;\zeta ^{2}=2m\mathcal{E}-k^{2}-\left(
gbB_{0}\right) ^{2},  \label{eq:dif_eq_G_s}
\end{eqnarray}
where $l=0,\pm 1,\pm 2,\ldots $ is the rotational quantum number, $s=\pm 1$,
$m$ the mass of the particle, $g$ a constant, $b$ the magnitude of the fixed
space-like vector field and $B_{0}$ the magnitude of the magnetic field. The
constant $-\infty <k<\infty $ comes from the fact that the motion is
unbounded along the $z$ axis; therefore the spectrum is continuous an
bounded from below $\mathcal{E}\geq \frac{\zeta ^{2}}{2m}+\frac{\left(
gbB_{0}\right) ^{2}}{2m}$. The authors simply set $\hbar =1$, $c=1$ though
there are well known procedures for obtaining suitable dimensionless
equations in a clearer and more rigorous way\cite{F20}.

In what follows we focus on the discrete values of $W$ that one obtains from
the bound-state solutions of equation (\ref{eq:dif_eq_G_s}) that satisfy
\begin{equation}
\int_{0}^{\infty }\left| G_{s}(\xi )\right| ^{2}\xi \,d\xi <\infty .
\label{eq:bound_states}
\end{equation}
Since the behaviour of the solution $G_{s}(\xi )$ at origin and at infinity
is determined by the terms $\nu _{s}^{2}\xi ^{-2}$ and $\xi ^{2}$,
respectively, we expect to have bound states for all $-\infty <\beta <\infty
$. Besides, the eigenvalues $W$ satisfy
\begin{equation}
\frac{\partial W}{\partial \beta }=\left\langle \frac{1}{\xi }\right\rangle
>0,  \label{eq:HFT}
\end{equation}
according to the Hellmann-Feynman theorem\cite{F39}.

By means of the ansatz
\begin{equation}
G_{s}(\xi )=\xi ^{\gamma }e^{-\frac{\xi ^{2}}{2}}P(\xi ),\;P(\xi
)=\sum_{j=0}^{\infty }a_{j}\xi ^{j},\;\gamma =\left| \nu _{s}\right| ,
\label{eq:ansatz}
\end{equation}
one obtains the three-term recurrence relation
\begin{eqnarray}
a_{j+2} &=&\frac{\beta }{(j+2)(2\gamma +j+2)}a_{j+1}+\frac{2j-\kappa }{%
(j+2)(2\gamma +j+2)}a_{j},  \nonumber \\
j &=&-1,0,1,2\ldots ,\;a_{-1}=0,\;a_{0}=1,  \label{eq:rec_rel}
\end{eqnarray}
where $\kappa =W-2\gamma -2$.

If the truncation condition $a_{n+1}=a_{n+2}=0$, $a_{n}\neq 0$ has
physically acceptable solutions then one obtains some exact eigenvalues and
eigenfunctions. The reason is that $a_{j}=0$ for all $j>n$ and the factor $%
P(\xi )$ in equation (\ref{eq:ansatz}) becomes a polynomial of degree $n$.
This truncation condition is equivalent to $\kappa =2n$ and $a_{n+1}=0$. The
latter equation is a polynomial function of $\beta $ of degree $n+1$ and it
can be proved that all the roots $\beta _{\gamma }^{(n,i)}$, $i=1,2,\ldots
,n+1$, $\beta _{\gamma }^{(n,i)}>\beta _{\gamma }^{(n,i+1)}$, are real\cite
{CDW00,AF20}. If $V(\beta ,\xi )=\beta /\xi +\xi ^{2}$ denotes the
parameter-dependent potential for the model discussed here, then it is clear
that the truncation condition produces an eigenvalue $W_{\gamma
}^{(n)}=2(n+\gamma +1)$ that is common to $n+1$ different potential-energy
functions $V_{\gamma }^{(n,i)}(\xi )=V\left( \beta _{\gamma }^{(n,i)},\xi
\right) $. It is worth noticing that the truncation condition only yields
\textit{some particular} eigenvalues and eigenfunctions because not all the
solutions $G_{s}(\xi )$ satisfying equation (\ref{eq:bound_states}) have
polynomial factors $P(\xi )$. From now on we will refer to them as
\begin{equation}
G_{\gamma }^{(n,i)}(\xi )=\xi ^{\gamma }P_{\gamma }^{(n,i)}(\xi )\exp \left(
-\frac{\xi ^{2}}{2}\right) ,\;P_{\gamma }^{(n,i)}(\xi
)=\sum_{j=0}^{n}a_{j,\gamma }^{(n,i)}\xi ^{j}.  \label{eq:G^(n,i)(xi)}
\end{equation}

Let us consider the first cases as illustrative examples. When $n=0$ we have
$\beta _{\gamma }^{(0)}=0$ and the eigenfunction $G_{\gamma }^{(0)}(\xi )$
has no nodes. We may consider this case trivial because the problem reduces
to the exactly solvable harmonic oscillator. Probably, for this reason it
was not explicitly considered by Bakke and Belich\cite{BB12}.

When $n=1$ there are two real roots
\begin{eqnarray}
\beta _{\gamma }^{(1,1)} &=&\sqrt{2}\sqrt{2\gamma +1},\;a_{1,\gamma
}^{(1,1)}=\frac{\sqrt{2}}{\sqrt{2\gamma +1}},  \nonumber \\
\beta _{\gamma }^{(1,2)} &=&-\sqrt{2}\sqrt{2\gamma +1},\;a_{1,\gamma
}^{(1,2)}=-\frac{\sqrt{2}}{\sqrt{2\gamma +1}},  \label{eq:beta,a,n=1}
\end{eqnarray}
and we appreciate that $G_{\gamma }^{(1,1)}(\xi )$ is nodeless while $%
G_{\gamma }^{(1,2)}(\xi )$ exhibits one node in $0<\xi <\infty $.

For $n=2$ we have
\begin{eqnarray}
\beta _{\gamma }^{(2,1)} &=&2\sqrt{4\gamma +3},\;a_{1,\gamma }^{(2,1)}=\frac{%
2\sqrt{4\gamma +3}}{2\gamma +1},\;a_{2,\gamma }^{(2,1)}=\frac{2}{2\gamma +1},
\nonumber \\
\beta _{\gamma }^{(2,2)} &=&0,\;a_{1,\gamma }^{(2,2)}=0,\;a_{2,\gamma
}^{(2,2)}=-\frac{1}{\gamma +1},  \nonumber \\
\beta _{\gamma }^{(2,3)} &=&-2\sqrt{4\gamma +3},\;a_{1,\gamma }^{(2,3)}=-%
\frac{2\sqrt{4\gamma +3}}{2\gamma +1},\;a_{2,\gamma }^{(2,3)}=\frac{2}{%
2\gamma +1}.  \label{eq:beta,a,n=2}
\end{eqnarray}
Notice that $G_{\gamma }^{(2,1)}(y)$, $G_{\gamma }^{(2,2)}(y)$ and $%
G_{\gamma }^{(2,3)}(y)$ have zero, one and two nodes, respectively, in the
interval $0<\xi <\infty $. It is also worth noticing that Bakke and Belich%
\cite{BB12} overlook the multiple roots mentioned above, as well as their
meaning.

In present notation the energy of the quantum-mechanical system reads
\begin{eqnarray}
\mathcal{E}_{\gamma }^{(n,i)} &=&\omega _{\gamma }^{(n,i)}\left( \gamma
+n+1\right) +\frac{\left( gbB_{0}\right) ^{2}}{2m}+\frac{k^{2}}{2m},
\nonumber \\
\omega _{\gamma }^{(n,i)} &=&\frac{\delta ^{2}}{m\left[ \beta _{\gamma
}^{(n,i)}\right] ^{2}},  \label{eq:E^(n,i)}
\end{eqnarray}
and is more general that the expression for $\mathcal{E}_{n,l,s}$ of Bakke
and Belich that does not take into account the possibility of multiple roots
of $a_{n+1}=0$. Since $\left[ \beta _{\gamma }^{(1,1)}\right] ^{2}=\left[
\beta _{\gamma }^{(1,2)}\right] ^{2}=2(2\gamma +1)$ and $\left[ \beta
_{\gamma }^{(2,1)}\right] ^{2}=\left[ \beta _{\gamma }^{(2,3)}\right]
^{2}=4(4\gamma +3)$ then our general expression (\ref{eq:E^(n,i)}) yields
the values of $\omega $ shown by these authors for $n=1$ and $n=2$. They
state that ``where the angular frequency of the harmonic oscillator is now
given by $\omega =\omega _{n,l,s}$ due to the condition $a_{n+1}=0$. The
condition $a_{n+1}=0$ allows us to obtain a expression involving the angular
frequency $\omega $ and the quantum numbers $n$, $l$ and $s$. By writing $%
\omega =\omega _{n,l,s}$, we have that the choice of the values of $\omega $
depends on the quantum numbers $n$, $l$ and $s$ in order to satisfy the
condition $a_{n+1}=0$. Thereby, we are assuming that $\omega _{n,l,s}$ can
be adjusted in order that the condition $a_{n+1}=0$ can be satisfied.''
Besides, in the conclusions the authors also state that ``Moreover, we have
shown that Lorentz symmetry breaking effects do not break the degeneracy of
the energy levels of the harmonic oscillator, but impose a dependence of the
cyclotron frequency on the quantum numbers $n$, $l$ and $s$.'' The main
point of present Comment is that the dependence of the oscillator frequency
on $n$, $l$ and $s$ is just an artifact of the truncation condition that
does not yield the spectrum of a given quantum-mechanical model but some
particular eigenvalues for different models. For example, $\mathcal{E}%
_{n,l,s}$ is an eigenvalue of some model $V_{\gamma }^{(n,i)}(\xi )$ while $%
\mathcal{E}_{n^{\prime },l^{\prime },s^{\prime }}$ is an eigenvalue of a
different quantum-mechanical model $V_{\gamma ^{\prime }}^{(n^{\prime
},i^{\prime })}(\xi )$; therefore, any conclusion derived from such
analytical expressions is meaningless from a physical (and, of course,
mathematical) point of view. Besides, it is clear that $n$ is not a quantum
number as shown by the case $n=2$ discussed above for which we obtained
three eigenfunctions with zero, one and two nodes and different values of $%
\beta $ (which, obviously, mean different model potentials).

In what follows we show how to do the calculation properly. For a given set
of values of the model parameters $m$, $g$, $b$, $B_{0}$ and the quantum
numbers $l$ and $s$ we obtain $\gamma $ and $\beta $. Then we solve the
eigenvalue equation (\ref{eq:dif_eq_G_s}) and obtain the allowed values of $W
$ for which the eigenfunctions are square integrable. If we call these
eigenvalues $W_{j,\gamma }$, $j=0,1,\ldots $, $W_{j,\gamma }<W_{j+1,\gamma }$
then the true energies of the system are given by
\begin{equation}
\mathcal{E}_{j,\gamma }=\frac{\omega W_{j,\gamma }}{2}+\frac{\left(
gbB_{0}\right) ^{2}}{2m}+\frac{k^{2}}{2m}.  \label{eq:E_(j,gamma)}
\end{equation}
Since the eigenvalue equation (\ref{eq:dif_eq_G_s}) is not exactly solvable
(it is in fact quasi solvable, or conditionally solvable\cite
{CDW00,AF20,F20b,F20c}, see also Turbiner's remarkable review\cite{T16}) we
resort to standard approximate methods; for example, the reliable Ritz
variational method that is known to yield upper bounds to all the
eigenvalues of a quantum mechanical model\cite{P68}. For simplicity, we
choose the basis set of non-orthogonal Gaussian functions $\left\{ u_{j}(\xi
)=\xi ^{\gamma +j}e^{-\frac{\xi ^{2}}{2}},\;j=0,1,\ldots \right\} $ and
reduce the problem to a secular equation that can be solved in a
straightforward way\cite{P68}. Since the eigenfunctions of the harmonic
oscillator $V(0,\xi )$ are linear combinations of these Gaussian functions
then present basis set is complete. In order to test the accuracy of the
variational results we resort to the powerful Riccati-Pad\'{e} method that
exhibits exponential convergence rate\cite{FMT89a}.

As a first illustrative example we choose $\gamma =0$ and $\beta =\sqrt{2}$
that comes from the truncation condition. The first four eigenvalues are: $%
W_{0,0}=W_{0}^{(1)}=4$, $W_{1,0}=7.693978891$, $W_{2,0}=11.50604238$, $%
W_{3,0}=15.37592718$. Notice that the truncation condition only yields the
ground state and misses all the other bound states. If, on the other hand,
we choose $\gamma =0$ and $\beta =-\sqrt{2}$ (the other root for $n=1$) the
first four eigenvalues are: $W_{0,0}=-1.459587134$, $W_{1,0}=W_{0}^{(1)}=4$,
$W_{2,0}=8.344349427$, $W_{3,0}=12.53290130$. In this case the truncation
condition misses all the spectrum but the first excited state.

As stated above, the eigenvalue equation (\ref{eq:dif_eq_G_s}) supports
bound states for all $-\infty <\beta <\infty $. In the first two examples we
chose $\beta =\pm \sqrt{2}$ that stem from the truncation condition. In what
follows we consider two $\beta $ values that are not roots of such
condition. When $\beta =-1$ and $\gamma =0$ the numerical approaches yield $%
W_{0,0}=-0.2085695649$, $W_{1,0}=4.601041510$, $W_{2,0}=8.834509671$, $%
W_{3,0}=12.96513798$. For $\beta =1$ and $\gamma =0$ we have $%
W_{0,0}=3.496523196$, $W_{1,0}=7.236061810$, $W_{2,0}=11.08720729$, $%
W_{3,0}=14.98768617$. If we consider the sequence $\beta =-\sqrt{2}%
,\;-1,\;1,\;\sqrt{2}$ we realize that each eigenvalue $W_{j,0}$ increases as
predicted by the Hellmann-Feynman theorem (\ref{eq:HFT}).

\textit{Summarizing}: since the eigenvalue equation (\ref{eq:dif_eq_G_s})
supports bound states for all $-\infty <\beta <\infty $ we conclude that
there are bound states for all values of $\omega >0$. The allowed values of
the oscillator frequency predicted by the truncation method are merely an
artifact of this approach and do not exhibit any physical meaning
whatsoever. The main basic and conceptual error in the authors' reasoning is
to believe that the only bound states are those with polynomial functions (%
\ref{eq:G^(n,i)(xi)}). The truncation method based on a three-term
recurrence relation only yields one and only one pair
eigenvalue-eigenfunction of a particular quantum-mechanical model.
The spectrum $\mathcal{E}_{n,l,s}$ shown by the authors is a
collection of eigenvalues of different quantum-mechanical systems.
The origin of this misunderstanding, which has spread over several
papers and several years, is a paper by Ver\c{c}in\cite{V91} in
which he states that there may be no bound states except for some
discrete values of the intensity of the magnetic field. However,
in a sequel Myrheim et al\cite{MHV92}  proved such conclusion
wrong and stated that ``Without the Coulomb interaction the odd
terms in the power series vanish, and the wave function is
normalizable if and only if the power series terminates. However,
when the even and odd terms are coupled by the Coulomb term ...,
there may exist infinite series solutions with physically
acceptable behaviour.'' They already calculated some of such
solutions and their figure 2 clearly shows that the eigenvalue of
the radial equation (proportional to $\nu $) is a continuous
function of the magnetic-field intensity (embedded in the
parameter $b$). Unfortunately, Myrheim et al\cite{MHV92} did not
explicitly state that Ver\c{c}in\'{}s conjecture was wrong and
several authors based their research on such false conclusion.

\end{document}